\documentclass{elsart}

\usepackage{epsfig}

\usepackage{amssymb}

\begin{document}
\newcommand{\be}{\begin{equation}}
\newcommand{\ee}{\end{equation}}

\begin{frontmatter}



\title{Modeling skin effect in large magnetized iron detectors}


\author[roma,lnf]{M.~Incurvati},\author[lnf]{~F.~Terranova\corauthref{cor} } 
\corauth[cor]{Corresponding author.}
\address[roma]{Dep. of Electric Engineering, Univ. 
``La Sapienza'', Roma, Italy}
\address[lnf]{Laboratori
Nazionali di Frascati dell'INFN, Frascati (Roma), Italy}

\begin{abstract}
The experimental problem of the calibration of magnetic field 
in large iron detectors is discussed. Emphasis is laid on techniques
based on ballistic measurements as the ones employed by MINOS or OPERA.
In particular, we provide analytical formulas to model the
behavior of the apparatus in the transient regime, keeping into account
eddy current effects and the finite penetration velocity of the driving fields.
These formulas ease substantially the design of the calibration apparatus.
Results are compared with experimental 
data coming from a prototype of the OPERA spectrometer.
\end{abstract}

\end{frontmatter}

\begin{figure}
\centerline{\epsfig
{file=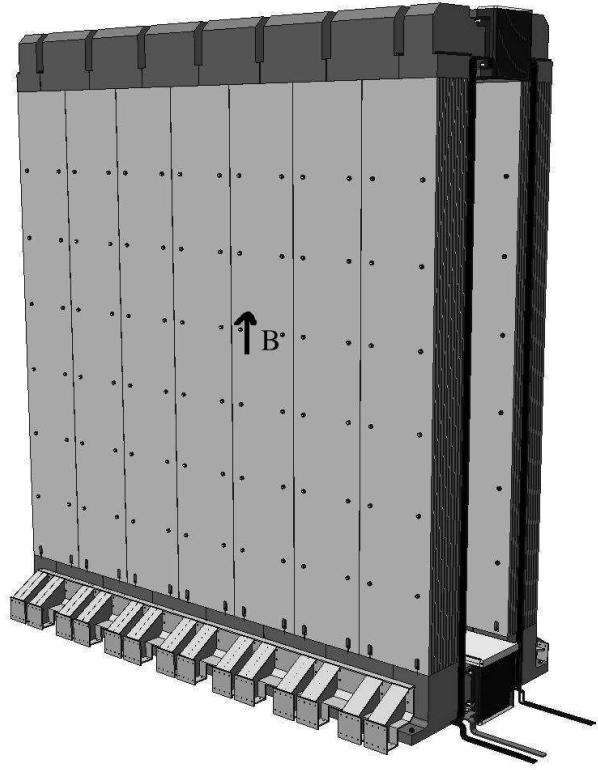,width=14cm} }
\caption{Schematic view of the OPERA spectrometer.}
\label{spectro} \end{figure}

Massive magnetized iron detectors will play a leading role in neutrino physics
in the forthcoming years. These detectors are used as instrumented target
(MINOS \cite{MINOS}) or complementing high precision trackers 
(OPERA \cite{OPERA}) in the next generation long baseline neutrino
experiments. Moreover, similar devices have been proposed for high
precision atmospheric neutrino experiments (MONOLITH \cite{MONOLITH})
or in connection with the Neutrino Factories \cite{NUFACT}. In most of their
physical applications a calibration of the magnetic field in the bulk of 
the iron is mandatory. In particular, for MINOS or OPERA a relative precision
on the knowledge in the local field of a few percent is enough to reach
a negligible impact on the charge and momentum reconstruction of muons.   
Absolute calibration by stopping muons requires long data taking
in the underground areas where these detectors will be located and will
not provide locally the requested precision. On the other hand, Hall probes
in the surrounding air test the field in iron
only indirectly
and need a detailed simulation to be rescaled. In fact, an absolute 
calibration of the average field in a given area 
can be obtained performing ``ballistic measurements'' \cite{zijlstra},
i.e. integrating the induced voltage in a set of pickup coils
during the ramp-up of the driving current. This technique is currently 
employed  by MINOS and OPERA \cite{meas_minos,meas_opera}. 
In particular, the OPERA dipolar magnet 
consists of two vertical walls of rectangular cross section
and of top and bottom flux return path (see Fig.~\ref{spectro}). 
The walls consist of 12 iron layers
(5~cm think) interleaved with 2~cm of air allocated for the housing of the
Resistive Plate Chambers. 
Each iron layer is made up of seven plates 
$50 \times 1250 \times 8200$~mm$^3$. The driving coils will be installed in the
return yokes. Several pickup coils (not shown in the figure) will be positioned
along the walls, measuring the field in iron at a given 
height averaged over the 12 slabs. 
During the ramp-up, the variation of the current flowing in the drive coils
induces a change in the magnetic flux cut by the pickup coil; 
the induced voltage is

\be
V(t)= -\frac{\d \Phi(t)}{\d t} = -\frac{\d}{\d t} 
\int_S \vec{B}(t)\cdot \vec{n} \ \d s = - SN \frac{\d}{\d t} 
\langle B \rangle (t)  
\label{fnl}
\ee

\noindent 
where $S$ is the cross sectional area (in iron) 
of one turn of the pickup coil, 
$N$ is the number of turns and $ \langle  B \rangle $ 
is the field averaged on the surface $S$.
For a ramp-like waveform of the current flowing in the driving coils

\begin{equation}
i(t) = \left\{ \begin{array}{r@{\quad:\quad}l}
0 & t<0 \\
kt & 0<t<T \\
kT & t>T   \end{array} \right. 
\label{waveform}
\end{equation}  

\noindent after integration of (\ref{fnl}), we get

\be
\frac{-1}{SN} \int_0^\infty 
\d t' V_{i}(t')  = \frac{1}{SN} ( \Phi_{\mathrm{fin}}-\Phi_0 )  =
\langle B \rangle_{\mathrm{fin}} - \langle B \rangle_0
\label{bave}
\ee

where the subscript  ``fin'' indicates the value reached
for $t \rightarrow +\infty $, i.e. when the power supply provides
a current $i=i_{\mathrm{max}}\equiv kT$ and the transient due to eddy currents
is faded out.   
Clearly, this measurement provides a difference of fields.
To get $B$ at the nominal value of $i_{\mathrm{max}}$ it is 
necessary to compute 
the residual field $B_r$; it can be done running through a
whole hysteresis loop, i.e. ramping up and down twice, and reconstructing
$B_r$ and $B(i_{\mathrm{max}})$ from the differences of fields.

The measurement based on Eq.~(\ref{bave}) is robust since
it does not depend on the actual waveform of the induced voltage 
(if the integration time is sufficiently long) and even the requirement
of a constant $di/dt$ can be safely dropped.
Hence, this technique can be extremely effective if $k$ ($\equiv \d i/\d t$) 
and $N$ are properly tuned, i.e. 
the induced voltage is enough and has a favorable
signal-to-noise ratio against environmental noise. 
In particular, the tuning of $k$ is critical and requires a careful design
of the power supplies connected with the driving coils
since the load inductance of the
spectrometer can be huge\footnote{The load inductance depends
on the magnetic permeability
and, hence, varies during the ramp-up. In the case of OPERA
it has a value  of about 0.14~H at nominal fields ($i=i_{\mathrm{max}}$) 
but during the ramp-up
reaches a maximum of 0.9~H.}. Although the field measurement
is independent of the waveform of the induced voltage, an approximate estimate
of this voltage is needed to design the apparatus and provides
useful information to assess systematic errors during the data taking. 
Neglecting eddy currents, $V(t)$ has a simple form:

\begin{eqnarray}
V(t)   =  -SN \frac{\d}{\d t} ( \mu(H) \cdot H ) =  
-SN \left[ \frac{\d \mu}{\d H} \frac{\d H}{\d t}
H(t) \ + \ \mu(H)\frac{\d H}{\d t} \right]  \nonumber  
\end{eqnarray}
\begin{eqnarray}
= \  -SNnk \left[\frac{\d \mu}{\d H}
H(t) \ + \ \mu(H) \right]  
\label{waveform_noskin}
\end{eqnarray}

\noindent being $H$ the magnetic intensity, 
$\mu(H)$ the magnetic permeability of iron and $n$ the effective number of
driving turns per unit length ($B=\mu H = \mu n i$).
For very simple geometries $n$ can be computed analytically. 
E.g. approximating the OPERA spectrometer
as an Epstein frame \cite{ASTM}, 
$n$ would be the number of driving coils
divided by the average path along the magnetic circuit. However, since
the driving coils are not uniformly distributed along the path
and the field losses in air are not negligible, the effective value
of $n$ at the height where the pickup coils are installed
has to be computed numerically by finite-element calculation
in the steady regime\footnote{In the present case
TOSCA \cite{TOSCA} has been used to solve numerically the magnetostatic Maxwell
equation for the actual geometry of the magnet.}. 
Once $n$ is obtained, Eq.~(\ref{waveform_noskin}) can be computed analytically
since $\mu(H)$ and its derivatives are known from the B-H curve of the iron
and $\d H/\d t=nk$ during the ramp-up. 
Eq.~(\ref{waveform_noskin}) provides also
a useful scaling law for different geometries. For instance, both the walls 
and the return yoke of the prototype
of the OPERA spectrometer \cite{meas_opera} 
are made up of 4 slabs instead of 12
but the magnetic properties of the steel are the same. The ratio
between the maximum voltage obtained during the ramp-up will be just

\be
\frac{V_{\mathrm{max}}}{V_{\mathrm{max}}'}=\frac{SNnk}{S'N'n'k'}  
\label{ratio}
\ee

where the primed quantities refer to the prototype, the others
to the final magnet. To prove Eq.~(\ref{ratio}) 
is enough noticing that $V$ is maximized when 
$f(H) \equiv (\d \mu/\d H) \cdot H + \mu$ 
reaches its maximum and $\mathrm{Max}\{f(H)\}$ is an universal 
property of the steel so it cancels out in the ratio (\ref{ratio}).

\begin{figure}
\centerline{\epsfig
{file=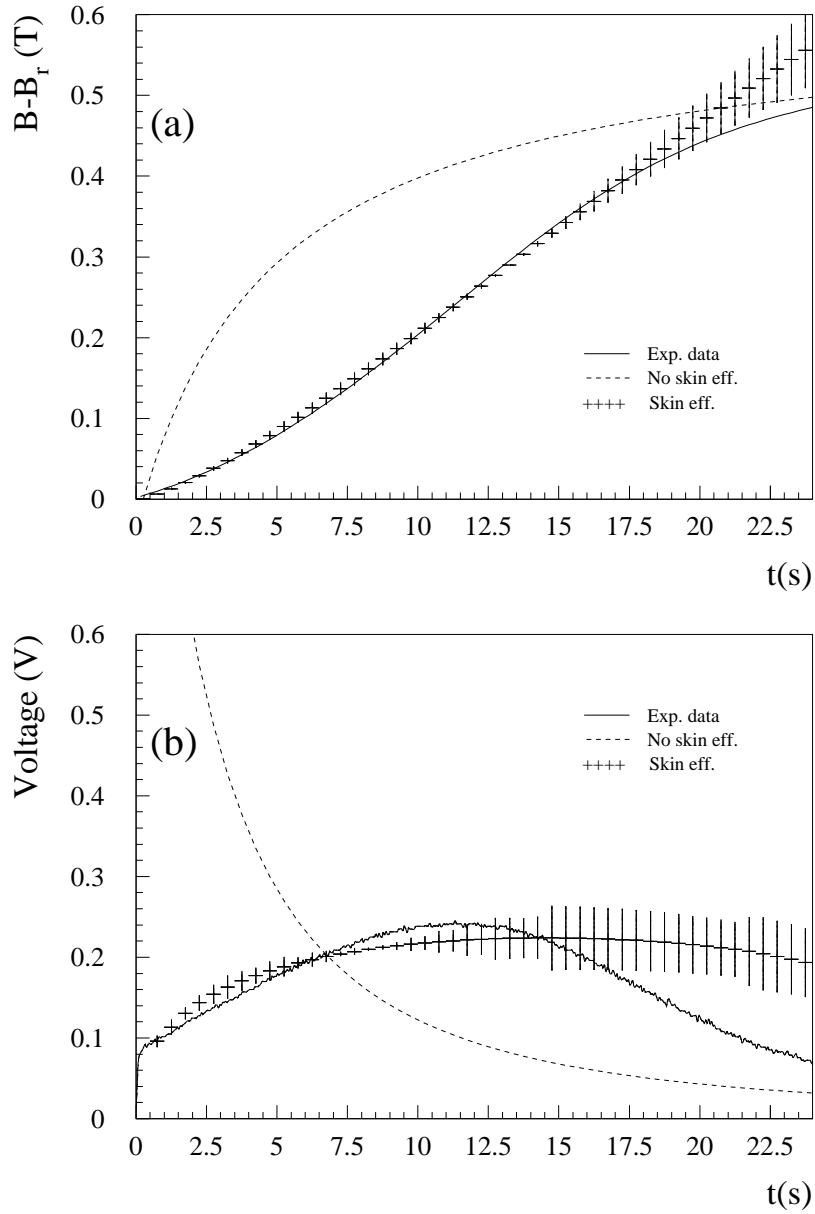,width=12cm} }
\caption{(a) Average magnetic field versus time measured
experimentally (continuous line), computed without skin effect (dashed line)
and computed including magnetic skin effect (crosses). (b) 
Induced voltage versus time   measured
experimentally (continuous line), computed without skin effect 
(dot-dashed line) and computed including magnetic skin effect (dots).}
\label{fields} \end{figure}

Unfortunately 
Eqs.(\ref{waveform_noskin}) and (\ref{ratio}) are completely 
spoiled by the magnetic
skin effect. Eddy currents induced in the bulk of the iron slow down
the penetration of the field through the slabs; hence the assumption
that $H(t)$ is equal to $nkt$ at any depth becomes unrealistic. 
To appreciate the relevance of this effect, 
Fig.~\ref{fields}-a shows the value of the field at one pickup coil
of the OPERA prototype
during a ramp-up from the residual field $B_r$ ($i=0$) up to $B^{\mathrm{max}}$
($i=368$~A). The dashed line represents the expected field without
eddy currents (i.e. assuming infinite resistivity for the steel),
the continuous line shows the experimental data 
($\rho=1 \cdot 10^{-7}$~$\mathrm{\Omega}$m$^{-1}$). Modeling the build up
of eddy currents during the transient regime through finite-element
calculation for these kind of devices is extremely cumbersome
and analytic or semi-analytic formulas to reproduce the magnetic behavior
of the spectrometer for time-varying currents would be of great practical
importance. Moreover, in the case of OPERA or MINOS, the problems can be 
reduced to the calculation of the penetration rate of a time varying field
$H_x(t,y=0)=H_x(t,y=D)=nkt$ 
through a semi-infinite plane of depth $D$ \footnote{In the following 
we use rectangular coordinates where the $x$- and $z$-directions
are parallel to the surface of the slab; the $y$-direction is into the slab,
normal to the surface. The interfaces between iron and air are at
$y=0$ and $y=D$. The air between the slabs is neglected. Note also
that in the present case the field $H$ is always parallel to $x$:
$\vec{H}(t)=H_x(t) \ \vec{i}$.}.
Here $D$ is the sum of the thicknesses of the slabs in the wall or in the
return yoke: $D=4\cdot 5$~cm~$=20$~cm for the OPERA prototype.      

This problem can be solved analytically for materials of constant permeability
\cite{bozorth}. For ferromagnetic materials several approximate 
solutions have been proposed in literature. In particular
it has been shown in \cite{bowden} that if the B-H curve of the steel
can be represented by the function

\be
B=aH^b \ \ \ 0<b<1
\label{param}
\ee

and the driving field is sinusoidal with frequency $\omega$
($H_{x0}=\hat{H}_{x0}  \cos  {\omega t}$),
the complete solution for $H_x$ in the slab is

\be
H_x(y,t)= \left\{ \begin{array}{l@{\quad:\quad}l}
 \hat{H}_{x0} (1-\alpha y)^\beta \cos { \left[ \omega t + 
\gamma \ ln(1-\alpha y) \right] } & \alpha y \le 1 \\
0 & \mathrm{elsewhere}  \end{array} \right. 
\label{solution}
\ee

\noindent where 

\be
\alpha \equiv \left[ \frac{ (1-b)^2 |\omega| \mu}
 {(3+b)\sqrt{2(1+b)} \rho} \right]^{1/2} \ \ ; \ \ 
\mu = a \hat{H}_{x0}^{b-1} 
\ee
\be
\beta \equiv \frac{2}{1-b} \ \ ; 
\ \ \gamma \equiv \frac{\sqrt{2(1+b)}}{(1-b)} 
\ee

\noindent The condition $\alpha y<1$ has a simple physical interpretation.
When $y \rightarrow 1/\alpha$, $H$ goes to zero, so that no electromagnetic 
fields exist beyond $1/\alpha$. Hence, every frequency
penetrates up to a given depth and it is confined within a ``skin''
of thickness $\alpha^{-1}$.  
In many applications, the power law (\ref{param}) is troublesome
because it is impossible to parametrize the whole
range of the B-H curve  with a single pair of constants.
In particular, the steepness $b$  changes
abruptly above the saturation knee when the rate of variation of $B$ 
becomes much smaller.
This means that during the ramp-up Eq.~(\ref{param}) remains valid
up to a time $\tilde{T}$ when the average field in the slabs
starts saturating. Beyond this point 
Eq.~(\ref{solution}) systematically overestimates
the field, due to the failure of (\ref{param}), and the model is no more
effective. Fortunately
this limitation is immaterial in the present case since we are
interested in reproducing the induced voltage in the region 
contributing mostly to the integral (\ref{bave}), i.e. below
the saturation knee. Moreover, since we need to describe $V(t)$ just
up to $\tilde{T}$, it is possible to choose a waveform different 
from (\ref{waveform}),   
simplifying the treatment in the Fourier domain. For instance,
we choose a triangular waveform

\begin{equation}
i(t) = \left\{ \begin{array}{r@{\quad:\quad}l}
0 & t<-\tilde{T} \\
k\tilde{T}+kt & -\tilde{T}<t<0 \\
k\tilde{T}-kt & 0<t<\tilde{T} \\
0 & t>\tilde{T}   \end{array} \right. 
\label{waveform2}
\end{equation}  

\noindent which is identical to (\ref{waveform}) in the ramping-up
phase. In this case, the time-varying field at $y=0$ 
( $H(t,y=0)=ni(t)$ ) has a non-singular real Fourier
transform:

\be
\frac{2\tilde{H}}{\tilde{T}\omega^2} (1-\cos { \omega \tilde{T}} )
\label{fourier}
\ee

\noindent where $\tilde{H}\equiv nk\tilde{T}$ is the maximum
field at the surface. 
\noindent The field at a given depth $d$ is

\be
H^{\mathrm{tot}}_x(t,y=d)=H_x(t,y=d)+H_x(t,y=D-d)
\ee

\noindent where

\be
H_x(t,y)= \frac{1}{2\pi} \int_{\mathcal{D}} \d \omega \ 
\frac{2\tilde{H}}{\tilde{T}\omega^2} (1-\cos {\omega \tilde{T} } )
(1-\alpha y)^\beta \cos {\left[ \omega t + 
\gamma ln(1-\alpha y) \right] } 
\ee

\noindent The integration domain $\mathcal{D}$ 
is given by the subset of the real
axis where the condition $\alpha y<1$ holds.
This integral has been computed numerically in the case of the OPERA
prototype. The crosses of Fig.~\ref{fields}-a represent
the expected value of the field according to the model described above. The
error band shows the systematic uncertainty coming from parametrization
(\ref{param}) and has been computed comparing (\ref{param}) with the actual
B-H~curve of the steel. Other source of errors as the border effects
coming from the finite length of the plane or the neglect of the air
between the wall slabs have been computed by finite element calculation
and are  negligible compared with the uncertainty coming from the 
parametrization. Fig.~\ref{fields}-b shows the corresponding induced
voltage versus time. Similar results have been obtained at other
hysteresis curves. Below the B-H~knee the model is able to reproduce
the experimental data within the instrumental precision (about 3\%
\cite{meas_opera}). For high values of $\tilde{H}$ the systematic bias
is within the parametrization uncertainty discussed above.

In conclusion, in this letter we provided a semi-analytic treatment
of the magnetic behavior of large magnetized spectrometers
in the transient regime. This model can be fruitfully applied
to design field calibration systems based on ballistic measurements,
especially for underground detectors where absolute
calibration from the range-curvature correlation of penetrating particles
is not available.

\begin{ack}
We are greatly indebted with R.~Rabinovici for drawing our attention to 
Ref.~\cite{bowden}. We thank J.~Nelson, M.~Spinetti and L.~Votano
for useful discussions on the ballistic measurements.
\end{ack}



\end{document}